\documentclass{iau} 
\usepackage{graphicx}

\newcommand\fs{\hbox{$.\!\!^{\rm s}$}}

\newcommand\farcs{\hbox{$.\!\!^{\prime\prime}$}}
\newcommand*\aap{A\&A}

\newcommand*\apj{ApJ}
\newcommand*\apjl{ApJ}

\newcommand*\apss{Ap\&SS}

\newcommand*\mnras{MNRAS}

\title[Phase connected X-ray light curve and He II RV measurements of NGC~300~X-1] 
{Phase connected X-ray light curve and He II radial velocity measurements of NGC~300~X-1}

\author[S. Carpano, F. Haberl, P. Crowther \& A. Pollock]   
{S. Carpano$^1$
 \and F. Haberl$^1$
 \and P. Crowther$^2$
 \and A. Pollock$^2$
 }

\affiliation{$^1$Max-Planck-Institut f\"{u}r extraterrestrische Physik, \\Giessenbachstra{\ss}e 1, 
85748 Garching, Germany\\ emails: {\tt scarpano@mpe.mpg.de, fwh@mpe.mpg.de} \\[\affilskip]
$^2$Department of Physics and Astronomy \& Space Physics, \\ University of Sheffield, Sheffield S3 7RH, UK
\\emails: {\tt paul.crowther@sheffield.ac.uk, a.m.pollock@sheffield.ac.uk}}

\pubyear{2019}
\volume{xxx}  
\jname{IAUS 346 High-mass X-ray binaries: illuminating the 
passage from massive binaries to merging compact objects}
\editors{A.C. Editor, B.D. Editor \& C.E. Editor, eds.}
\begin{document}

\maketitle

\begin{abstract}
NGC~300~X-1 and IC~10~X-1 are currently the only two robust extragalactic candidates for being Wolf-Rayet/black hole X-ray binaries, the Galactic analogue 
being Cyg~X-3. These systems are believed to be a late product of high-mass X-ray binary evolution and direct progenitors of black hole mergers. From the analysis 
of Swift data, the orbital period of NGC 300 X-1 was found to be 32.8\,h. We here merge the full set of existing data of NGC~300~X-1, using XMM-Newton, 
Chandra and Swift observations to derive a  more precise value of the orbital  period of 32.7932$\pm$0.0029~h above a confidence level of 99.99\%. This allows us to phase 
connect the X-ray light curve of the source with radial velocity measurements of He~II lines performed in 2010. We show that, as for IC~10~X-1 and Cyg~X-3, the X-ray 
eclipse corresponds to maximum of the blueshift of the He~II lines, instead of the expected zero velocity.
This indicates that for NGC~300~X-1 as well, the wind of the WR star is completely ionised by the black hole radiation  and that the emission lines come from the region of
the WR star that is in the shadow. We also present  for the first time the light curve of two recent very long XMM-Newton observations of the source, performed on the 16th to 20th of December 2016.

\end{abstract}

\firstsection 
\section{Introduction}
Wolf-Rayet/black hole X-ray binaries are believed to be a late product of high-mass X-ray binaries and direct progenitors of black hole mergers (\cite[Bulik et al. 2011, Bogomazov 2014]{Bulik2011, Bogomazov2014}). 
Following the standard model, when the massive secondary star completes its main sequence evolution, it becomes a supergiant filling its Roche Lobe. The system is then embedded in a common envelope and the binary components may get closer. The envelope from the massive star is lost leaving the system composed of a Wolf-Rayet star and a relativistic object (here a black hole). 

The association of the X-ray source NGC 300 X-1 with a Wolf-Rayet star was reported by \cite[Carpano et al. (2007a) and Crowther et al. (2007)]{Carpano2007a,Crowther2007}. From the analysis of Swift data, the orbital periods for NGC~300~X-1 and IC~10~X-1,  were found to be 32.8\,h \cite[(Carpano et al. 2007a)]{Carpano2007b} and 34.9\,h (\cite[Prestwich et al. 2007]{Prestwich2007}), respectively, and were confirmed in the optical band by radial velocity (RV) measurements of He~II lines \cite[(Crowther et al. 2010, Silverman  \& Filippenko 2008)]{Crowther2010,Silverman2008}. 

A more precise value of 34.84306\,h for the period of IC~10~X-1 has been provided by \cite[Laycock et al. (2015a)]{Laycock2015a}, by combining XMM-Newton and Chandra data. This allowed to phase connect the X-ray light curve and the RV data showing that the X-ray eclipse is associated to the maximum of the blueshift of the He II line, instead of the expected zero velocity, found for example for the X-ray binary M33~X-7 \cite[(Orosz et al. 2007)]{Orosz2007}. This feature was also reported for the Galactic Wolf-Rayet/black-hole X-ray candidate Cyg~X-3 \cite[(van~Kerkwijk 1993, van~Kerkwijk et al. 1996, Hanson et al. 2000)]{vanKerkwijk1993, vanKerkwijk1996, Hanson2000}. The model invoked to explain this 1/4 phase shift, is that the wind of the Wolf-Rayet star is completely ionised by the black hole radiation, except in the region shadowed by the star from which most of the emission lines are originating. As explained in more detailed in \cite[Laycock et al. (2015b)]{Laycock2015b} for IC10~X-1, during X-ray eclipse when the WR star lies between the black hole and the observer, the wind velocity vector lies along our line of sight producing the maximum blueshift of the emission lines. At a quarter of phase earlier or later, the wind in the shielded  sector is orthogonal to the line of sight, producing zero Doppler shift.


\section{Observations and data reduction}
\subsection{Swift observations}
For this work we used 86 Swift XRT observations from the 5th of September 2006 to 19th June 2018 performed in photon counting mode and with total exposures varying from $\sim$300\,s to $\sim$10000\,s. The first long set was triggered on December 2016, after the discovery that NGC~300~X-1 was associated with a Wolf-Rayet star, and led to the discovery of the 32.8\,h orbital period. 

The event files were barycenter-corrected using the \texttt{ftool barycorr} from the HEASoft package.  For every observation, we extracted the number of source and background counts obtained in every GTI defined in the file header. Net count rates were then obtained by subtracting the number of background counts from the source counts after scaling for the respective areas and by dividing by the GTI length.

\subsection{XMM-Newton observations}
There are currently 7 XMM-Newton observations available for NGC 300, performed between December 2000 and December 2016. The first four, published in \cite[Carpano et al. (2005), Carpano~et al. (2007a)]{Carpano2005,Carpano2007a}, have an exposure of about $\sim$40\,ks each, while the fifth, pointing on a supernova, was lasting only $\sim$20\,ks. The last two observations performed on three consecutive days have a very long exposure of 140+80\,ks, to cover two orbital periods. These were carried out almost simultaneously to a 163\,ks long NuSTAR observation, starting on the 16th of December 2016, to study the hard (3--79\,keV) component of the electromagnetic spectrum.

The data were reduced following standard procedures using the XMM-Newton SAS data software version 15.0.0.  Single to quadruple events were used for MOS (\cite[Turner~et al. 2001]{Turner2001}) cameras, with FLAG=0, while NGC~300~X-1 was partially or totally in a pn (\cite[Str{\" u}der~et al. 2001]{Strueder2001}) CCD gap for all observations. The event files were barycenter-corrected using the SAS \texttt{barycen} tool.

\subsection{Chandra observations}
There are currently 5 Chandra observations available with NGC~300~X-1 in the field-of-view, 4 of them with the Advanced CCD Imaging Spectrometer (ACIS) instrument (\cite[Garmire et al. 2003]{Garmire2003})  and one with the High Resolution Camera (\cite[HRC, Murray et al. 2000]{Murray2000}), in the period between June 2006 and November 2014, with exposures from 10 to 65\,ks. The data were reduced using the Chandra X-ray data analysis software, \texttt{ciao} version 4.9. The event files were barycenter-corrected using the \texttt{axbary} tool.

\section{Light curve analysis}
\subsection{Refinement of the orbital period}
\begin{figure}[!]
   \centering
   \resizebox{\hsize}{!}{\includegraphics{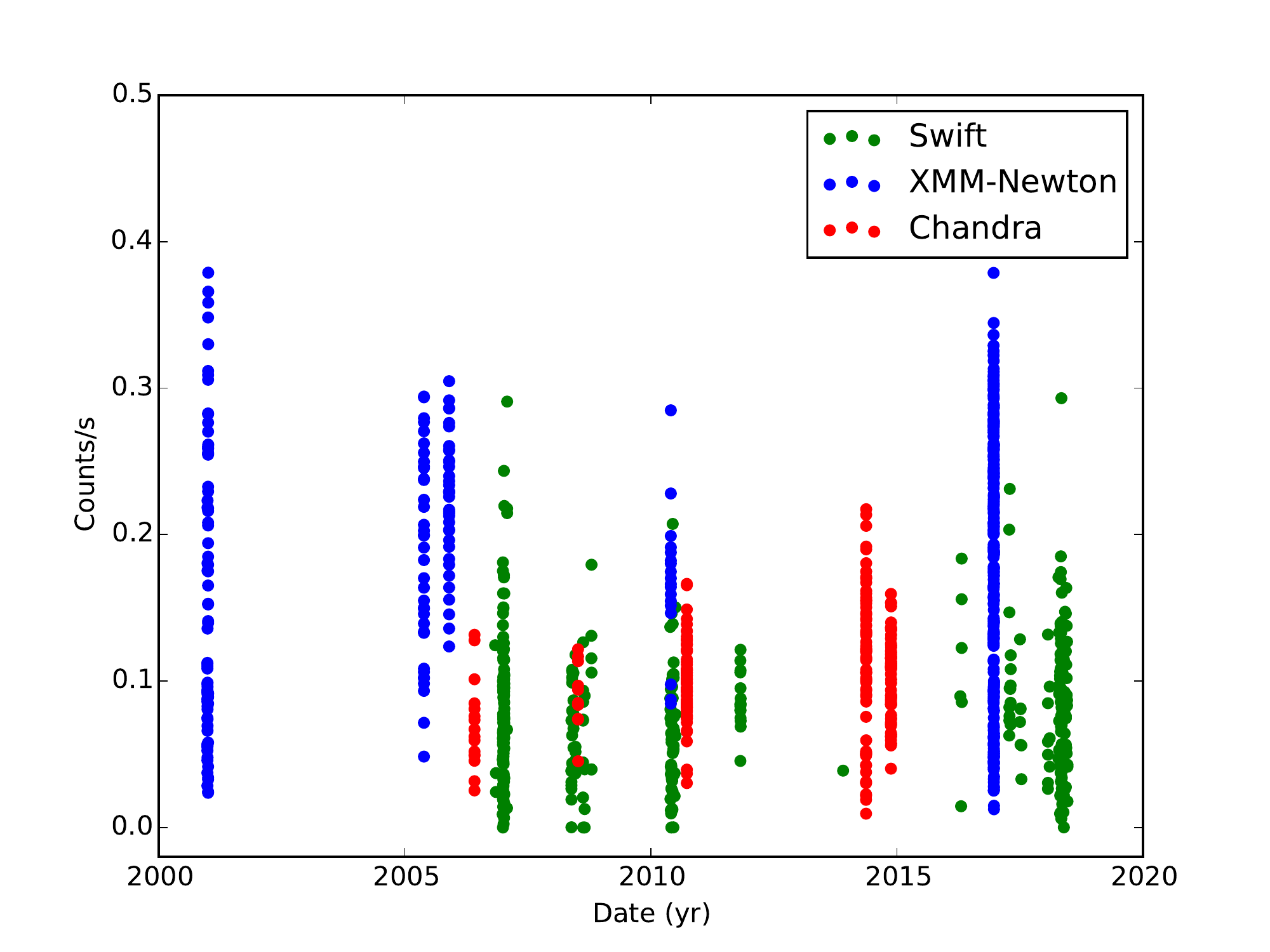}}
   \caption{Long-term light curve of NGC~300~X-1, containing the Swift (green), XMM-Newton (blue) and Chandra (red) data and taking into account differences in the effective area from the various instruments and vignetting issues. The count rate values represent what is expected for MOS1+MOS2 data.}
              \label{fig:long_lc}             
\end{figure}

We refined the value of the orbital period by combining all available observations from Swift, XMM-Newton and Chandra. The source extraction region was centered around the Chandra coordinates derived by \cite[Binder et al. (2011)]{Binder2011}, which is $\alpha_\textrm{J2000}=00^\textrm{h}55^\textrm{m} 10\fs{}00$, $\delta_\textrm{J2000}=-37^\circ 42' 12\farcs 2$.   For the XMM-Newton observations, we only merged MOS1 and MOS2 data, since the source lies partially or totally in a pn CCD gap for all observations. We take into account differences in the effective area from the various instruments and vignetting issues. For this, we created simulated source spectra for every observation using the \texttt{fakeit} command of XSPEC with the relevant response files. The model spectrum is an absorbed power-law (\texttt{phabs*power}) with an $N_\textrm{H}$ value of 5$\times$10$^{20}$\,cm$^{-2}$ and  $\Gamma$=2.45, similar to the model fitted to the first 4 XMM-Newton observations in \cite[Carpano~et al. (2007a)]{Carpano2007a}. The count rate of the simulated spectra, extracted in the 0.2--10.0\,keV band, is then used to scale the observed count rates, since these represent the values expected for a constant source. The light curve segments from XMM-Newton and Chandra are rebinned to 1000\,s, while for Swift, there is a data point for every valid GTI. The long-term light curve,  spanning over more than 17 years, is shown in Fig.~\ref{fig:long_lc} with a count rate value scaled to what is expected for MOS1+MOS2.

\begin{figure}[!]
   \centering
   \resizebox{\hsize}{!}{\includegraphics{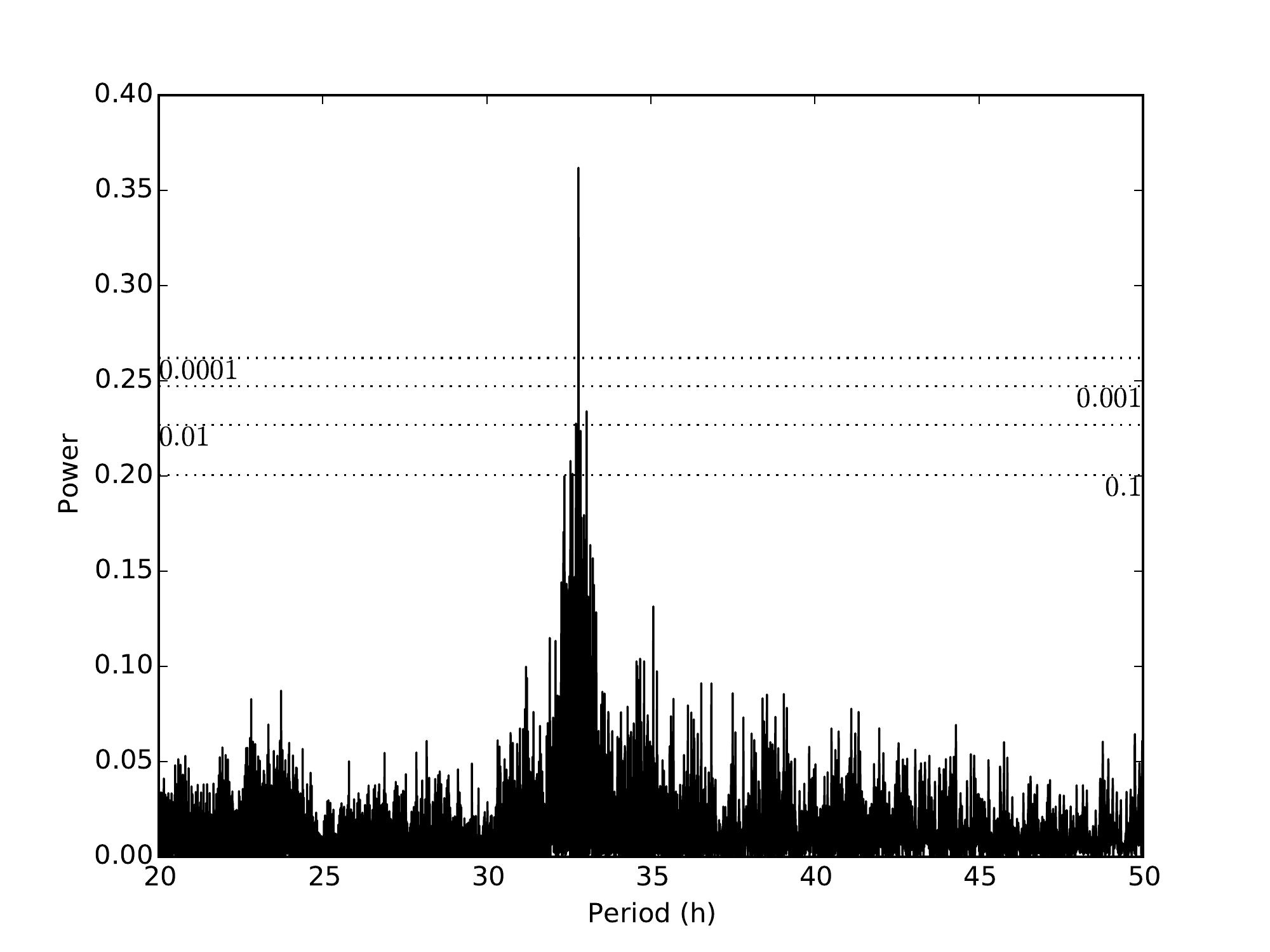}}
   \caption{Lomb-Scargle periodogram in the 20 to 50\,h, from the merged Swift, XMM-Newton and Chandra light curve shown in Fig.~\ref{fig:long_lc}, with confidence levels at 90\%, 99\%, 99.9\% and 99.99\%, assuming white noise.}
              \label{fig:period}             
\end{figure}

We then determined the new value of the orbital period using the Lomb-Scargle periodogram analysis \cite[(Lomb~1976, Scargle~1982)]{Lomb1976,Scargle1982}, in the full (0.2--10\,keV) energy band and in the period range from 20 to 50\,h, using the Python algorithm from the \texttt{gatspy} library. Fig.~\ref{fig:period} shows the result of the period search, with a high peak at 32.7932\,h above a confidence level of 99.99\%, measured using Monte Carlo simulations and assuming white noise. We estimate the period error to be 0.0029\,h by associating it to the 1$\sigma$ width of a gaussian function fitted on the highest peak.

\subsection{Phase connected folded X-ray light curve and RV measurements}
\begin{figure}[!]
   \centering
   \resizebox{\hsize}{!}{\includegraphics{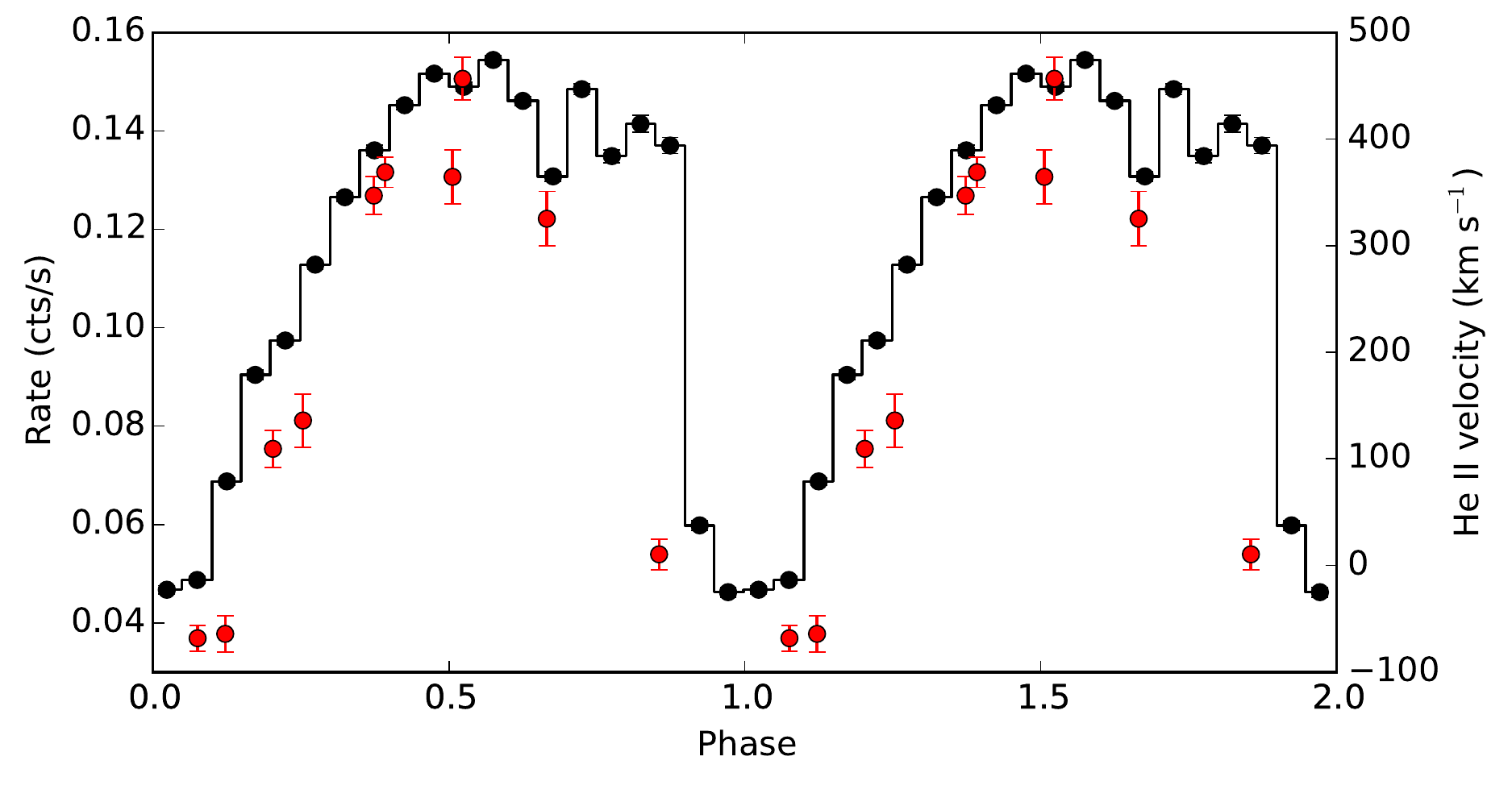}}
   \caption{X-ray light curve of NGC~300~X-1 of Fig.~\ref{fig:long_lc} with the VLT/FORS He II radial velocity measurements (red) folded at the best period ($\sim$32.7932\,h). Phase 0 corresponds to the start of the first XMM-Newton observation.}
              \label{fig:fold_lc}             
\end{figure}

In Fig.~\ref{fig:fold_lc}, we fold the long term X-ray light curve of NGC~300~X-1 of Fig.~\ref{fig:long_lc} with the best period derived from the periodogram. We also overlay the radial velocity measurements of the VLT/FORS He II lines ($\lambda$4686) reported in Table~1 of \cite[Crowther~et al. (2010)]{Crowther2010} folded at the same period, after converting times to the same reference. Phase 0 corresponds to the start of the first XMM-Newton observation performed on the 26th of December 2000. As for IC~10~X-1 and Cyg~X-3, the X-ray eclipse corresponds to maximum of the blueshift of the He II lines, instead of the expected zero velocity.

\subsection{Analysis of the 2016 XMM-Newton observations}
\begin{figure}[!]
   \centering
   \resizebox{\hsize}{!}{\includegraphics{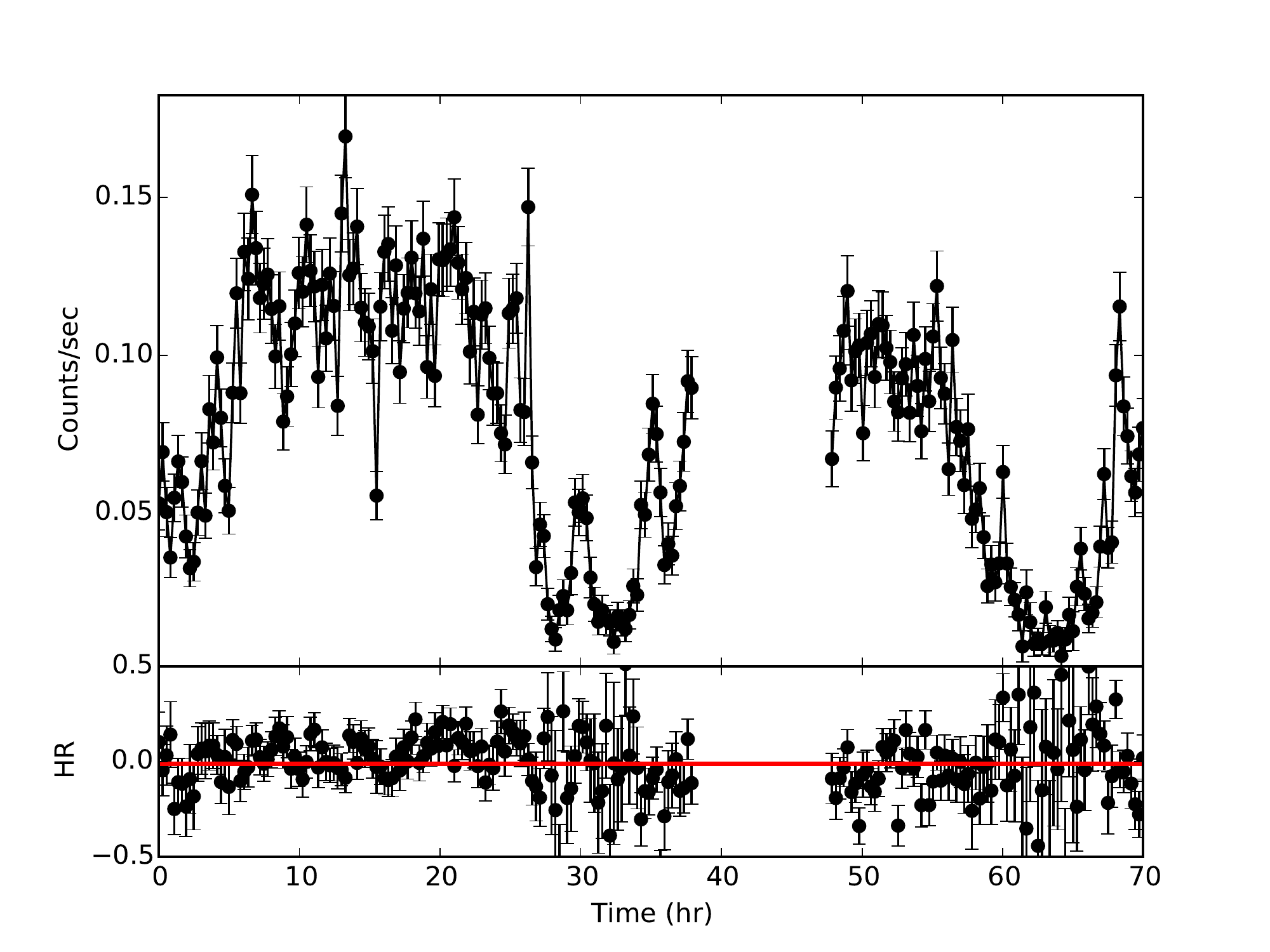}}
   \caption{Lightcurve from the 17th to 20th of December 2016 XMM-Newton observation (top) and hardness ratios HR=(H-S)/(H+S), with S=0.2--1\,keV and H=1--10\,keV.}
              \label{fig:2016_lc}             
\end{figure}

NGC~300~X-1 was observed during two long (135+80\,ks) observations from 17th to 20th of  December 2016, covering a bit more than two orbital cycles. The observation was performed simultaneously with NuSTAR, but the source was too soft to be detected in this harder 3--79\,keV band. Fig.~\ref{fig:2016_lc} (top) shows the corresponding light curve, expanding over about 70\,h with a gap of about 9\,h, and the hardness ratio (bottom) defined as HR=(H-S)/(H+S) with the soft and hard bands being 0.2--1\,keV and 1--10\,keV respectively. 

It is the first time the full orbit is covered continuously allowing a proper visualization of the light curve structure. We can notice that there is some erratic X-ray variability in the entire light curve likely caused by X-ray scattering of the photons by the wind of the WR star. There is no significant variation in the hardness ratios between inside and outside eclipse. We also note the presence of flares during the eclipse. 

\section{Conclusions}
We have shown in this work that one has to be cautious when measuring black hole or neutron star masses using radial velocity measurements. For all three Wolf-Rayet/compact object systems NGC~300~X-1, IC~10~X-1 and Cyg~X-3, a shift of 1/4 of orbital phase was observed between the folded radial velocity measurements and the X-ray light curves, using the He~II lines. This indicates that somehow the lines are not emitted isotropically and therefore do not trace the motion of the Wolf-Rayet star preventing to use them to calculate the compact object mass function. An explanation for this anisotropy could be that the Wolf-Rayet wind is fully ionised by the black hole radiation except in the part that is in the shadow.

One should therefore look for other absorption/emission lines that have the right phase in the RV curve with respect to the X-ray eclipse (zero velocity at minimum X-ray flux), in order to estimate the compact object mass. For Cyg~X-3, He I absorption lines, whose RV curve had the right phase but a lower amplitude, were detected during an observation when the system was in outburst. This helped to derive a more reliable value of the compact object mass function of 0.027\,M$_\odot$ (\cite[Hanson~et al. 2000]{Hanson2000}).

\begin{discussion}

\discuss{Andreas Sander}{Can you tell me anything about the Hydrogen content? Assuming that it is similar to Cyg~X-3, which is hydrogen-free,
one could assume that the moderate terminal velocity compared to single hydrogen-free WNs might be an argument for the wind being significantly affected by irradiation.}

\discuss{Stefania Carpano}{Indeed, like for Cyg~X-3, there are no Hydrogen lines in the spectrum of the Wolf-Rayet companion star of NGC~300~X-1.}
\end{discussion}

\end{document}